\documentclass[12pt,a4paper,final]{iopart}
\usepackage{iopams}  
\usepackage{units}
\usepackage{graphicx}
\usepackage{units}
\usepackage[breaklinks=false,colorlinks=true,linkcolor=blue,urlcolor=blue,citecolor=blue]{hyperref}

\def\dblone{\hbox{$1\hskip -1.2pt\vrule depth 0pt height 1.6ex width 0.7pt \vrule depth 0pt height 0.3pt width 0.12em$}}

\RequirePackage[xcolornames,svgnames,dvipsnames,rgb]{xcolor}
\pdfpageattr {/Group << /S /Transparency /I true /CS /DeviceRGB>>}

\newcommand\T{\rule{0pt}{2.6ex}}
\newcommand\B{\rule[-1.2ex]{0pt}{0pt}}

\begin{document}

\title[]{Tunnelling anisotropic magnetoresistance effect of single adatoms on a noncollinear magnetic surface}

\author{Nuala M. Caffrey, Silke Schr\"{o}der,  Paolo Ferriani and Stefan Heinze}
\address{Institut f\"ur Theoretische Physik und Astrophysik, 
Christian-Albrecht-Universit\"at zu Kiel, Leibnizstr. 15, 24098 Kiel, Germany}
\ead{caffrey@theo-physik.uni-kiel.de}

\begin{abstract}
The tunnelling anisotropic magnetoresistance (TAMR) effect describes the sensitivity of spin-polarized electron transport to the orientation of the magnetization with respect to the crystallographic axes.  As the TAMR effect requires only a single magnetic electrode, in contrast to the tunnelling magnetoresistance effect, it offers an attractive route towards alternative spintronics applications. 
In this work we consider the TAMR effect at the single-atom limit by investigating the anisotropy of the local density of states in the vacuum above transition-metal adatoms adsorbed on a noncollinear magnetic surface, the monolayer of Mn on W(110).
This surface presents a cycloidal spin spiral ground state with an angle of 173$^\circ$ between neighbouring spins and thus allows a quasi-continuous exploration of the angular dependence of the TAMR of adsorbed adatoms  using scanning tunnelling microscopy.
Using first-principles calculations, we investigate the TAMR of Co, Rh and Ir adatoms on Mn/W(110) and relate our results to magnetization direction dependent changes in the local density of states. The anisotropic effect is found to be enhanced dramatically on the adsorption of heavy transition-metal atoms, with values of up to 50\% predicted from our calculations. This effect will be measurable even with a non-magnetic STM tip. 
\end{abstract}


\section{Introduction}

Spin-dependent tunnelling between ferromagnetic electrodes separated by an insulating barrier is the basis of many magnetic data storage technologies. Many of these technologies are based on the tunnelling magnetoresistance (TMR) effect whereby the current that flows through such a device is strongly dependent on the relative magnetization of the magnetic layers due to the spin-dependent density of states. Switching between parallel (P) and antiparallel (AP) alignment causes a change in the device resistance \cite{Julliere1975225, parkin2004giant, lee2007effect}. 
TMR is generally defined as  (I$_\mathrm{P}$ - I$_\mathrm{AP}$) / I$_\mathrm{AP}$ where I$_\mathrm{P}$ is the current through the junction with parallel magnetic alignment and I$_\mathrm{AP}$  is the current through the junction with antiparallel alignment. 
Using scanning tunnelling microscopy (STM), where the role of the second electrode is played by a magnetic tip and tunnelling occurs through the vacuum rather than through an insulating layer, it is possible to detect a TMR effect for even single magnetic adatoms on 
surfaces \cite{PhysRevLett.99.067202, meier2008revealing, PhysRevLett.103.057202, Serrate2010, loth2010measurement, loth2010controlling, Ziegler2011, PhysRevB.86.180406, khajetoorians2013current}. 
Further to this, due to spin-orbit coupling, the resistance can also depend on the magnetization direction relative to the crystallographic axes, an effect which has been coined the tunnelling anisotropic magnetoresistance (TAMR) effect~\cite{PhysRevLett.93.117203}.
The TAMR effect is a particularly attractive route for alternative spintronics applications as only one ferromagnetic electrode is required and the effect does not require coherent spin-dependent transport.
\begin{figure}
\begin{centering}
\includegraphics[width=0.85\linewidth]{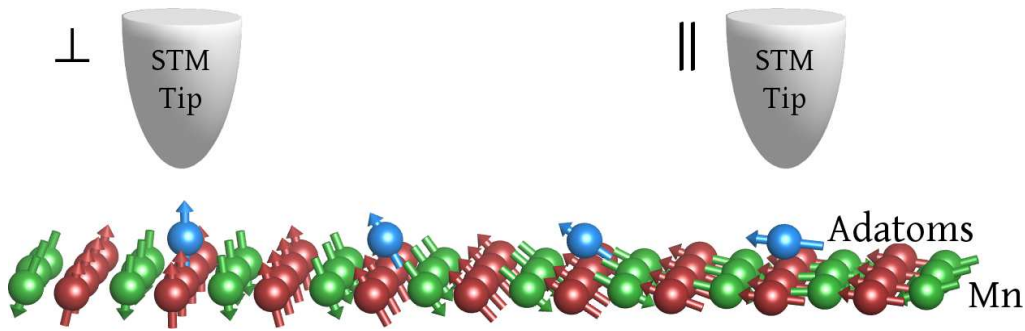}
\caption{\label{fig:tmr_tamr}(Color online) Schematic of a route to measure a TAMR effect above a single adatom coupled magnetically to an underlying substrate with a spin-spiral ground state using a non-magnetic STM tip. Here, the spin-spiral state in a Mn monolayer on the W(110) surface is used since it allows one to obtain out-of-plane (left), in-plane (right), as well as intermediate spin directions in a quasi-continuous manner. (W atoms not shown).}
\end{centering}
\end{figure}
The TAMR effect was first experimentally observed in $dI/dV$ measurements taken above the domains and domain walls of a double layer of Fe on W(110) where the magnetization pointed out-of-plane and in-plane, respectively \cite{PhysRevLett.89.237205}. 
TAMR effects were subsequently found in planar ferromagnetic junctions \cite{PhysRevB.73.024418, PhysRevLett.98.046601},  tunnel junctions \cite{PhysRevLett.93.117203, PhysRevB.79.155303, PhysRevB.80.045312, PhysRevLett.99.226602} and mechanically controlled break junctions \cite{Viret2006}. 
The TAMR effect also allows the possibility of imaging noncollinear spin structures on the atomic scale using STM with non-magnetic tips 
\cite{PhysRevB.86.134422}, although the measured corrugation amplitudes were quite small.
As the effect is driven by spin-orbit coupling, attempts have been made to increase its magnitude by combining 3$d$ and 5$d$ elements in bimetallic alloys \cite{PhysRevB.81.212409} and with antiferromagnetic electrodes \cite{park2011spin}.

A microscopic understanding of the mechanisms governing the TAMR effect, and in particular how it scales down to the atomic scale, can be achieved by considering the limiting case of a single adatom. Serrate \etal found evidence of spin-orbit related effects in Co adatoms adsorbed on a Mn monolayer (ML) grown on a W(110) substrate~\cite{Serrate2010}.
This surface displays a spin spiral ground state with spins propagating along the [1$\bar{1}$0] direction with an angle of $\approx$173$^\circ$ between neighbouring magnetic moments (see Fig.~\ref{fig:tmr_tamr}) \cite{Bode2007}.
In this sense it acts as an atomic-scale noncollinear magnetic template so that by moving adatoms laterally on the surface it is possible to control to a high degree the spin direction of the adatom. 
The Co adatom couples ferromagnetically to the underlying magnetic thin film via Heisenberg exchange, allowing for the spin to take any direction that is accessible in the Mn spin-spiral -  not only those directions perpendicular and parallel to the plane defined by the surface. STM line profiles along the [1$\bar{1}$0] direction above a row of deposited Co adatoms showed that although the height above the atomic centres followed an approximate $\cos \theta$ profile as expected for the TMR effect, where $\theta$ is the angle between the tip magnetization direction and the magnetization of the respective Co adatom, better agreement could be achieved by including a small percentage of the higher order correction $\cos^2 \theta$. It was proposed that this correction could be attributed to spin-orbit coupling, and in particular the TAMR effect.
More recently, N\'{e}el \etal  considered the change in electronic properties of a Co adatom adsorbed on the domain or domain wall of a double layer Fe film grown on the W(110) substrate \cite{Neel2013}. 
They performed spectroscopy measurements on Co adatoms with out-of-plane and in-plane magnetization direction and found a TAMR effect that repeatedly changes sign as a function of bias voltage and with a magnitude as large as 12\%. The observed TAMR was explained based on a first-principles calculation of the anisotropy of the local density of states in the vacuum. 

Here, we propose to study adatoms on noncollinear spin structures in order to explore the angular dependence and the magnitude of the TAMR. 
An advantage of using a surface with noncollinear magnetic ordering is that the magnetization direction of the adatom can be tuned almost 
continuously without requiring an external magnetic field.
This is achieved instead relying only on the local exchange interaction between the adatom and the surface. 
We use first-principles calculations based on density functional theory to investigate valence isoelectronic transition-metal adatoms of increasing atomic number, namely Co, Rh and Ir, on a Mn monolayer on W(110) which displays a spin-spiral ground state.
We consider the two limiting cases of the spin direction, i.e., when the magnetization direction points perpendicular to the surface and when it points in-plane along the [1$\bar{1}$0] direction, as shown in Fig.~\ref{fig:tmr_tamr}. 
We find that the adsorption of such adatoms can be used to enhance the spin-orbit coupling strength locally above a surface and in doing so offers a route to the determination of whether or not a surface spin structure displays noncollinearity using conventional STM with 
non-magnetic tips. In particular, the TAMR is dramatically enhanced for heavy 5$d$ transition-metal adatoms, which become spin-polarized by the substrate. 

This paper is organised as follows. After briefly presenting the computational method and the details of the calculation, we proceed to describe the electronic structure of the three adatoms adsorbed on the surface, focusing in particular on the vacuum density of states and the subsequent TAMR effect. The origin of the TAMR effect is explained within the framework of a simple model that considers the hybridization of two atomic states at a surface via spin-orbit coupling. The final section summarizes our main conclusions.

\section{Methods \& Computational Details}
In this work, density functional theory calculations are performed using the {\sc vasp} code \cite{Kresse1996, Kresse1999}. The Perdew-Burke-Ernzerhof (PBE) \cite{Perdew1996} parametrization of the generalized gradient approximation (GGA) is employed. The projector-augmented wave (PAW) method \cite{PhysRevB.50.17953} is used with the standard PAW potentials supplied with the VASP distribution. The plane wave basis set is converged using a 400~eV energy cutoff. Structural relaxations are carried out using a 6 $\times$ 6 $\times$ 1 k-point Monkhorst-Pack mesh \cite{PhysRevB.13.5188} to sample the three-dimensional Brillouin zone. Spin-orbit coupling is taken into account as described in Ref.~\cite{PhysRevB.62.11556}. Here, 576 k$_{\parallel}$-points are used in the full two-dimensional  Brillouin zone for the calculation of the local density of states (LDOS). 
The density of states in the vacuum was determined by positioning an empty sphere at the required height above the adatom onto which the DOS was projected. 

\begin{figure}
\begin{centering}
\includegraphics[width=0.65\linewidth]{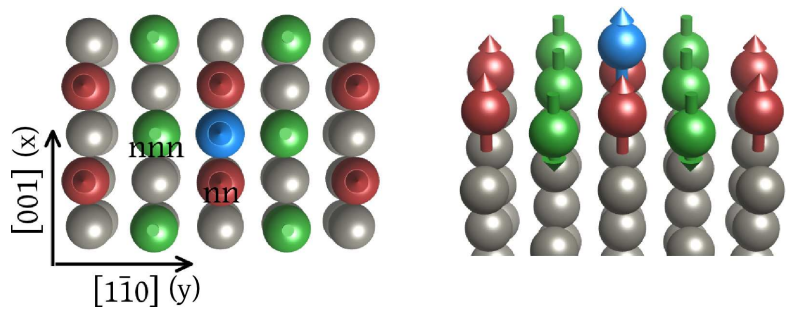}
\caption{\label{fig:structure}(Color online) Top and perspective view of the employed supercell. Grey spheres represent the W substrate atoms while the Mn atoms are depicted as red (green) spheres with arrows that are parallel (antiparallel) to the magnetic moment of the adatom (blue sphere). The nearest neighbour and next nearest neighbour Mn atoms to the adatom are labelled nn and nnn, respectively. Collinear antiferromagnetic ordering is assumed in all calculations. x and y refer to the notation used for the labeling of the $d$-orbitals. }
\end{centering}
\end{figure}

The system is modelled using a symmetric slab consisting of 5 atomic layers W with a pseudomorphic Mn layer on each side, as was found experimentally \cite{bode1999growth}.
Due to the long period of the spin spiral, the local magnetic order can be considered approximately collinear, especially when considering its interaction with localised adsorbed adatoms. We therefore approximate it as a two-dimensional antiferromagnet, using a $c$(2$\times$2) AFM surface unit cell, as shown in Fig.~\ref{fig:structure}. 
The GGA calculated lattice constant of W is 3.17~\AA, in good agreement with the experimental value of 3.165 \AA. The adatom has been added on each Mn surface in the hollow-site position \cite{Serrate2010} and the minimum distance between the adatoms in adjacent unit cells is 6.34~\AA\ so that any interaction between them will be negligibly small. Additionally, a thick vacuum layer of approximately 25~\AA\ is included in the direction normal to the surface to ensure no spurious interactions between repeating slabs. The positions of the Mn atoms as well as the adatom are optimized until all residual forces are less than 0.01~eV/\AA. In all cases the coordinates of the W atoms are held fixed and the relaxed interlayer distance between Mn and W is found to be 2.08~\AA.

\section{Results and discussion}

\subsection{Magnetic properties of adatoms on Mn/W(110)}

In this section, we consider how hybridization between the chosen adatoms and the surface modifies the magnetic properties of both. 
This is shown in Table~\ref{tab:magnetic_moments}. The magnetic moment of a Mn atom in the Mn monolayer on the W(110) surface is $\pm3.5$~$\mu_B$. For the case of a Co adatom, this moment is significantly reduced due to the adsorption of the adatom, as the Co $d$ states hybridize with $d_{xz}$ and $d_{z^2}$ states of the nearest neighbour Mn atoms (not shown). The largest drop in moment occurs for the nearest neighbour Mn atoms, where it is reduced by $0.9$~$\mu_B$. The Co adatom itself is found to have a magnetic moment of $+1.6$~$\mu_B$ orientated parallel to that the nearest neighbour Mn atoms beneath it when adsorbed on the surface in good agreement with Ref.~\cite{Serrate2010}. 
\begin{table}[ht]
\begin{centering}
\begin{tabular}{lccccc}%
\multicolumn{6}{c}{}\\
\hline \hline 
	\T \B 	&  	& Co &  & Rh & 	Ir	\\
\cline{1-6}
{\small Adatom}			\T \B		&	&	$+1.6$	&	
&	$+0.3$	&	$+0.2$\\
{\small Mn (nn)}	\T \B		&	&	$+2.6$	&	
&	$+2.9$	&	$+2.8$\\
{\small Mn (nnn)}		\T \B		&	&	$-3.3$	&	
&	$-3.4$	&	$-3.4$\\
{\small Mn (clean surface)} 	\T \B		&	&	$\pm3.5$	&	
&	$\pm3.5$	&	$\pm3.5$\\
\hline
\end{tabular}
\caption{\label{tab:magnetic_moments}\,Magnetic moments, in $\mu_B$, of the adsorbed Co, Rh and Ir adatoms and the Mn atoms in the underlying Mn monolayer on W(110)
(nn=nearest neighbour, nnn=next-nearest neighbour with respect to the adatom).
}
\end{centering}
\end{table}
This modification of the magnetic moments is accompanied by a slight buckling of the Mn monolayer in the vicinity of the adatom. This is shown in the first column of Table~\ref{tab:buckling}. The largest displacements occur for the next nearest neighbour Mn atoms, which move towards the adatom. 

Rh and Ir are valence isoelectronic to Co, but are non-magnetic. Upon adsorption on the Mn surface they develop a small magnetic moment due to their hybridization with the magnetic Mn atoms. For the case of Rh, the adatom has a small magnetic moment of $0.3$~$\mu_B$ after adsorption which is orientated parallel to magnetic moments of the row of Mn atoms beneath it along the [001] direction (See Table~\ref{tab:magnetic_moments}). The same is true for the case of the Ir adatom which develops a small magnetic moment of $0.2$~$\mu_B$. The buckling of the surface increases for the Rh and Ir adatoms with the next nearest neighbour Mn atoms moving towards the adatoms in all cases. 

\begin{table}[ht]
\begin{centering}
\begin{tabular}{lcccccc}%
\multicolumn{7}{c}{}\\
\hline \hline 
	\T \B 	&  	& Co &  & Rh & 	& Ir	\\
\cline{1-7}
{\small d (Adatom - $\:\,\,$nn Mn)}				\T \B		&	&	$\:\:\:\,2.26$	&	&	$\:\:\:\,2.38$	&	&	$\:\:\:\,2.37$\\
{\small d (Adatom - nnn Mn)}				\T \B		&	&	$\:\:\:\,2.62$	&	&	$\:\:\:\,2.65$	&	&	$\:\:\:\,2.59$\\
{\small $\Delta z_{\,\mathrm{nn}}$}		\T \B		&	&	$-0.07$	&	&	$-0.04$	&	&	$+0.05$\\
{\small $\Delta z_{\,\mathrm{nnn}}$}		\T \B		&	&	$+0.07$	&	&	$+0.12$	&	&	$+0.13$\\
{\small $\Delta x_{\,\mathrm{nn}}$}		\T \B		&	&	$+0.02$	&	&	$-0.05$	&	&	$-0.08$\\
{\small $\Delta y_{\,\mathrm{nnn}}$} 		\T \B		&	&	$+0.09$	&	&	$+0.12$	&	&	$+0.15$\\
\hline
\end{tabular}
\caption{\label{tab:buckling}\,Distance, in \AA, of Co, Rh and Ir adatoms from nearest neighbour (nn) and next nearest neighbour (nnn) Mn atoms of the underlying Mn monolayer on W(110). $\Delta x,y,z$ gives the displacement,  with respect to the clean surface, of the nn and nnn Mn atoms after adatom adsorption in \AA. Positive (negative) values imply the Mn atoms are moving towards (away from) the adatom.
}
\end{centering}
\end{table}

\begin{figure}[ht]
\begin{centering}
\includegraphics[width=0.45\linewidth]{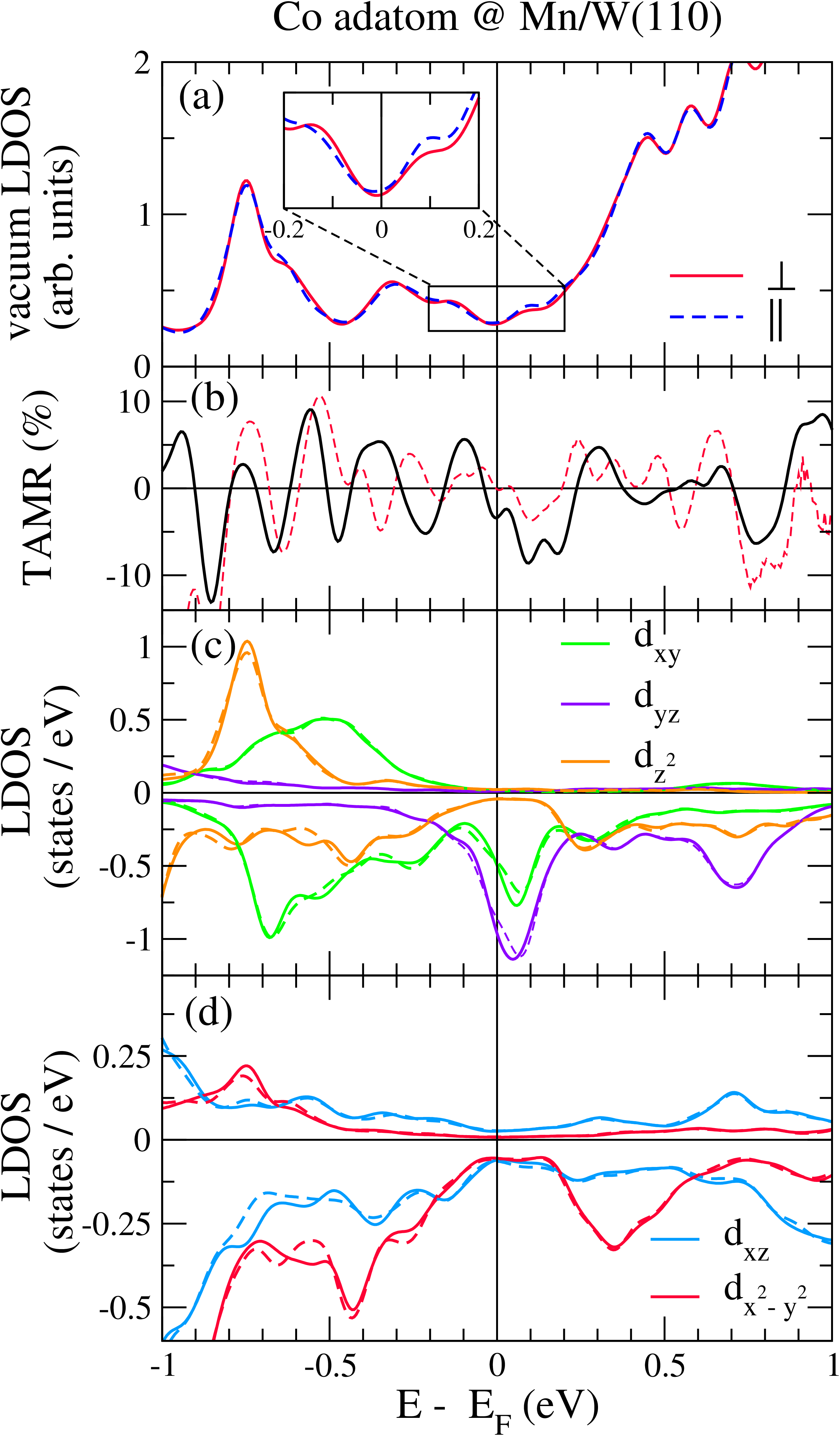}
\caption{\label{fig:co_adatom}(Color online)\,(a) LDOS in the vacuum evaluated at 6~\AA\ above the Co adatom for spin-quantization axes aligned parallel to the [1$\bar{1}$0] direction (blue, dashed) and perpendicular (red, solid) to the film plane. (b) TAMR effect of the data presented in (a) according to Equation~(\ref{eqn:tamr}). The red dashed line refers to the anisotropy of the LDOS at Co adatom considering only the $d_{z^2}$ states. (c) \& (d) LDOS of the Co adatom decomposed in terms of the orbital symmetry of the $d$-states. Solid (dashed) lines refer to the magnetization direction pointing perpendicular (parallel) to the film plane.}
\end{centering}
\end{figure}

\subsection{Co adatoms on Mn/W(110)}
We consider now the TAMR effect driven by the adsorption of a single magnetic Co adatom on the Mn ML on a W(110) surface. 
Using the spectroscopic mode of a STM, and within the Tersoff-Hamann model \cite{PhysRevB.31.805}, the differential conductivity ($dI/dV$) is directly proportional to the local density of states (LDOS) at the tip position, i.e.~in the vacuum a few \AA ngstr\"{o}m above the surface \cite{PhysRevLett.86.4132}. By measuring the differential conductivity above an adatom with two different directions of magnetization direction one can extract the TAMR, defined here as [($dI/dV$)$_\perp$ - $dI/dV$)$_\parallel$] / ($dI/dV$)$_\perp$. Therefore, we can theoretically evaluate the TAMR by calculating the anisotropy of the LDOS due to spin-orbit coupling.
Fig.~\ref{fig:co_adatom}(a) illustrates the differences that occur in the vacuum local 
density of states calculated above the Co adatom with different magnetization directions, i.e., aligned to the [1$\bar{1}$0] direction ($n_{\parallel}(z,\epsilon)$) and perpendicular to the surface ($n_{\perp}(z,\epsilon)$) at a height of $z=6$~\AA\ above the adatom. At first glance, both curves appear very similar. However, several difference emerge on a closer look.  The most significant features are located at $-0.75$~eV, $-0.1$~eV and $+0.1$~eV with respect to the Fermi energy (see inset of Fig.~\ref{fig:co_adatom}(a)). 
The anisotropy of the vacuum LDOS, i.e., the TAMR, is then depicted in Fig.~\ref{fig:co_adatom}(b), defined here as:
\begin{equation}\label{eqn:tamr}
 \mathrm{TAMR} = \frac{ n_{\perp}(z,\epsilon) - n_{\parallel}(z,\epsilon) }{n_{\perp}(z,\epsilon)}
\end{equation}
It exhibits values from $-16$\% to $+10$\% with some of the largest peaks occurring at the energies discussed previously. The large peak at $-0.9$~eV corresponds to an enhanced LDOS of the in-plane magnetised adatom compared to that of the out-of-plane magnetised adatom and results in a negative value of TAMR. 
Just below the Fermi level a significant enhancement occurs  for $n_{\perp}$ compared to $n_{||}$, whereas the reverse is true just above the Fermi level.
As the largest contribution to the LDOS in  the vacuum comes from states with $d_{z^2}$ symmetry (as well as $s$ and $p_z$), it follows that the largest contribution to the TAMR will come from a modification of the $d_{z^2}$ states. This is highlighted in Fig.~\ref{fig:co_adatom}(b) where the red dashed line shows the anisotropy of the Co $d_{z^2}$ states  at the atom. It is evident that the majority of the TAMR features in this energy range originate from changes in the $d_{z^2}$ states upon the change of magnetization direction due to the spin-orbit induced mixing of states. 
In order to determine these contributions the magnetization-direction dependent Co DOS is shown in Fig.~\ref{fig:co_adatom}(c) and (d). 
The majority states are almost entirely occupied so that the main contribution to the DOS at the Fermi level comes from the minority states, and in particular states with $d_{yz}$ and  $d_{xy}$ symmetry. These states decay relatively quickly in the vacuum, in contrast to the majority states which are mainly made up of the more slowly decaying $s$ and $d_{z^2}$ states. As a consequence, the vacuum DOS is comprised mainly of majority states and the pronounced peak exhibited by the minority spin states at the Co close to the Fermi level is absent entirely in the vacuum. This behaviour has been observed before both experimentally and theoretically \cite{Serrate2010}. 
The largest contribution to the TAMR at $-0.9$~eV originates from spin-orbit induced mixing of these $d_{z^2}$ states with states of  $d_{x^2- y^2}$ and $d_{xz}$ character whereas the peaks in the TAMR close to the Fermi level originate from interactions between the $d_{z^2}$ states and minority states of $d_{xz}$ and $d_{yz}$ symmetry. 

\subsubsection{Relating TAMR to the corrugation amplitude}
In an STM experiment the tunnelling current measured above adatoms with two different magnetization directions at a given bias voltage will be composed of a constant part, $I_\perp$ or $I_{\parallel}$, depending on the contributing electronic states and a height dependent part so that $I_\perp(z) = I_\perp e^{-2\kappa z_\perp}$ and $I_\parallel(z) = I_\parallel e^{-2\kappa z_\parallel}$.
In the limit of low bias voltages the decay constant can be written as $\kappa = \sqrt{2m\phi/\hbar^2}$ where $\phi$ is the apparent height of the tunnelling barrier.
If we now define the TAMR as $\frac{I_\perp - I_\parallel}{I_\perp}$ (i.e., using the current rather than the differential conductivity), we find that the difference in tip-height between in-plane and out-of-plane magnetization will be given as:
$$
\Delta z = z_\perp - z_\parallel = -\frac{1}{2\kappa}\ln(1-\frac{I_\perp - I_\parallel}{I_\perp})
$$
This implies that a TAMR of 10\% will give change in tip height of about $5$~pm between an adatom with an out-of-plane and an in-plane magnetization direction, the magnitude of which agrees well with the experimental data reported in Ref.~\cite{Serrate2010}. 

As an aside, we note here the implications of experimentally extracting the TAMR from the current measurements. As this method integrates all states within the bias window, it is possible that as the bias is increased the TAMR will become smaller, as any oscillations present in the asymmetry of the local density of states cancel each other. Such oscillations can be seen in Fig~\ref{fig:co_adatom}(b). As such, this method is most useful when very small voltages are applied, as in this regime the current will be approximately proportional to the local density of states in the vacuum \cite{PhysRevLett.86.4132, PhysRevB.83.214410}. For arbitrary voltages, however, the experimental TAMR is best determined using the spectroscopic mode, i.e., by measuring the differential conductivity and calculating the TAMR as [($dI/dV$)$_\perp$ - $dI/dV$)$_\parallel$] / ($dI/dV$)$_\perp$.  The differential conductivity will now be directly proportional to the local density of states and so an immediate comparison with the theoretically determined TAMR, as defined in Eqn.~\ref{eqn:tamr}, can be made. This is the approach taken in Ref.~\cite{PhysRevLett.89.237205} and Ref.~\cite{Neel2013}.

\subsection{Rh adatoms on Mn/W(110)}
\begin{figure}[ht]
\begin{centering}
\includegraphics[width=0.45\linewidth]{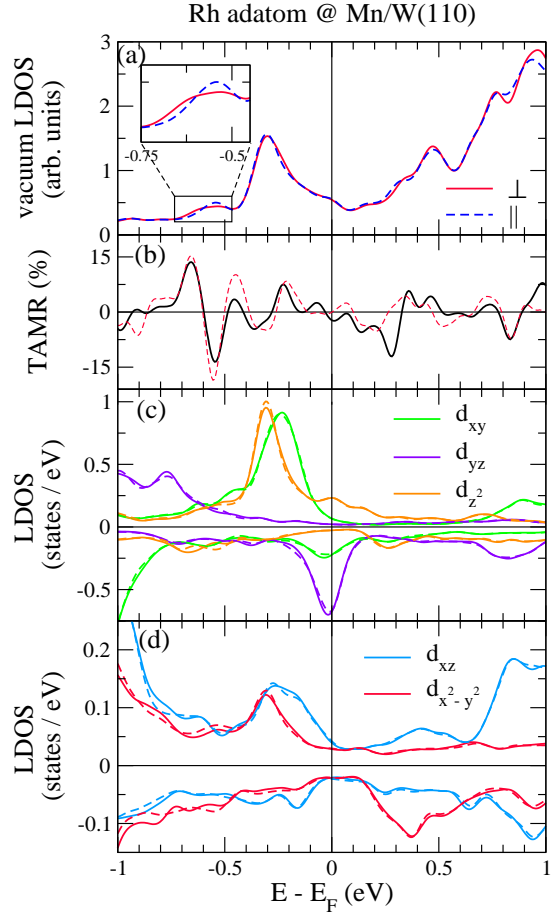}
\caption{\label{fig:rh_adatom}(Color online)\,(a) LDOS in the vacuum evaluated at 6~\AA\ above the Rh adatom on Mn/W(110) for spin-quantization axes aligned parallel to the [1$\bar{1}$0] direction (blue, dashed) and perpendicular (red, solid) to the film plane. (b) TAMR effect of the data presented in (a) according to Equation.~(\ref{eqn:tamr}).  The red dashed line refers to the anisotropy of the LDOS at Rh adatom considering only the $d_{z^2}$ states. (c) \& (d) LDOS of the Rh adatom decomposed in terms of the orbital symmetry of the $d$-states. Solid (dashed) lines refer to the magnetization direction pointing perpendicular (parallel) to the film plane.}
\end{centering}
\end{figure}

As we have seen, the SOC effect is quite weak in elements from the 3$d$ series. One method to increase the TAMR effect would be to adsorb a heavier element, from the 4$d$ or 5$d$ series, as the strength of the spin-orbit interaction in atoms increases quickly with its nuclear charge.
In this section and the next, we will consider the effect of a Rh ($Z = 45$) and an Ir ($Z = 77$) adatom
on the TAMR effect. Both atoms are valence isoelectronic to Co and non-magnetic but, as we have shown, develop a small magnetic moment when adsorbed on the surface due to a spin-dependent hybridization with the magnetic Mn atoms on the surface. The LDOS of the Rh $d$-states is plotted in Fig.~\ref{fig:rh_adatom}(c) and (d). The spin splitting of the Rh adatom
is significantly reduced compared to the magnetic Co atom with the majority $d_{z^2}$ states moving by $0.45$~eV closer to the Fermi level. The Fermi level itself is dominated by states with minority $d_{yz}$ character. Similar to the Co case, these minority states decay quickly in the vacuum so that by 6~\AA\ above the adatom states with majority $d_{z^2}$ character contribute more to the vacuum DOS. 
Fig.~\ref{fig:rh_adatom}(a) shows the vacuum density of states calculated above the Rh adatom with the two different magnetization directions, $n_{||}$ and $n_{\perp}$, again at a height of 6~\AA\ above the Rh adatom. The largest differences now occur at $-0.6$~eV and $+0.3$~eV where the former is created by changes in the minority $d_{z^2}$ states and the latter by changes in the majority $d_{z^2}$ LDOS. These differences correspond to a TAMR effect ranging from $+16$\% to $-15$\% (Fig.~\ref{fig:rh_adatom}(b)). Close to the Fermi level, the effect is slightly smaller ranging from $-10$\% to $+6$\% due to the smaller spin-orbit induced mixing at this energy level. 
Despite the higher atomic number of Rh compared to Co, it is evident that they both exhibit a similarly modest TAMR on this surface. This is, in part, due to the fact that Rh is a non-magnetic atom with small splitting between the orbitals that are interacting via spin-orbit coupling. The TAMR effect will be maximized for adatoms with large magnetic moments and large spin orbit coupling strengths.  While Rh, as a 4$d$ metal, has an increased spin orbit coupling strength compared to Co, the magnetic moment, and hence the spin splitting of the $d$-states, is significantly smaller (c.f. Table~\ref{tab:magnetic_moments}). The result is a slightly larger TAMR effect, but one that is relatively small considering its atomic weight. 

\subsection{Ir adatoms on Mn/W(110)}

\subsubsection{TAMR}

\begin{figure}[ht]
\begin{centering}
\includegraphics[width=0.45\linewidth]{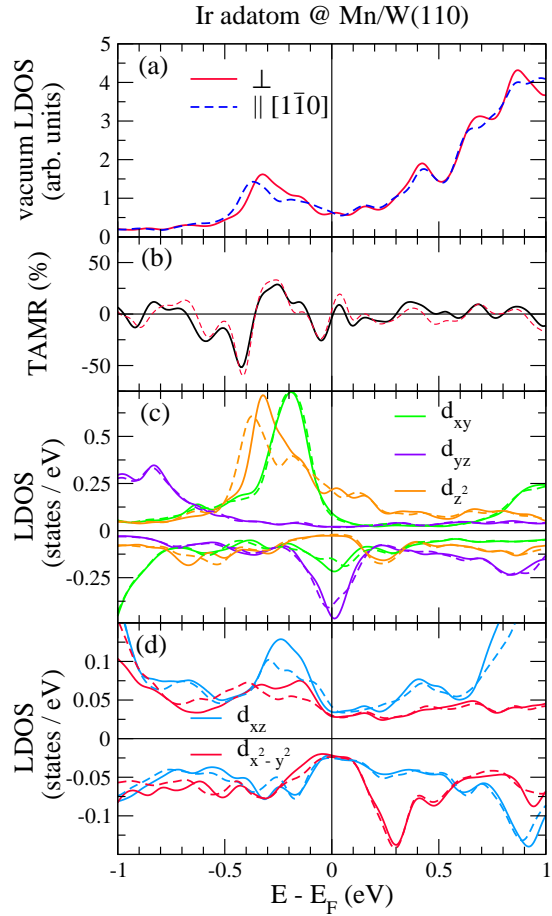}
\caption{\label{fig:ir_adatom}(Color online)\,(a) LDOS in the vacuum evaluated at 6~\AA\ above the Ir adatom for spin-quantization axes aligned parallel to the [1$\bar{1}$0] direction (blue, dashed) and perpendicular (red, solid) to the film plane. (b) TAMR effect of the data presented in (a) according to Equation.~(\ref{eqn:tamr}).  The red dashed line refers to the anisotropy of the LDOS at Ir adatom considering only the $d_{z^2}$ states. (c) \& (d) LDOS of the Ir adatom decomposed in terms of the orbital symmetry of the $d$-states. Solid (dashed) lines refer to the magnetization direction pointing perpendicular (parallel) to the film plane.}
\end{centering}
\end{figure}

Due to its large nuclear charge, and subsequent increased spin-orbit coupling strength, the Ir adatom exhibits a giant anisotropy of its LDOS. This results in a massive increase in the TAMR compared to the case for the 3$d$ and 4$d$ adatoms, in spite of a small induced magnetic moment of $+0.2$~$\mu_B$. Fig.~\ref{fig:ir_adatom}(a) shows the spin-averaged local density of states evaluated in the vacuum for the two different magnetization directions of the Ir adatom. 
In contrast to the very similar local density of states presented for the Co and Rh adatoms, here the curves for parallel and perpendicular magnetizations present many differences throughout a large range between $-1$~eV and $+1$~eV. The largest change is immediately visible at  $\approx -0.35$~eV and corresponds to a huge TAMR effect of $-50$\%. Closer to the Fermi level, at $-60$~meV, further difference in the two vacuum density of states can be seen, leading to a TAMR effect of $-30$\% just below the Fermi level and $+10\%$ just above it. 
A closer look at the orbitally resolved local density of states at the Ir adatom (Fig.~\ref{fig:ir_adatom}(c) and (d)) shows that the first peak in the TAMR stems mainly from a shift of the $d_{z^2}$ state to lower energies and broadening when the magnetization 
rotates from out-of-plane to in-plane along the [$1\bar{1}0$] direction. 
The TAMR effect at the Fermi level, is likewise due to magnetization-direction dependent changes in the $d_{z^2}$ state. 
Changes in the minority  $d_{yz}$ states can be seen at the same energy. This indicates that a spin-orbit induced hybridization between these two states generates the TAMR effect at this energy. 
Further evidence of the large contribution of the anisotropy of the $d_{z^2}$ states alone to the TAMR can be seen clearly in Fig.~\ref{fig:ir_adatom}(b) where both curves almost coincide. 
The essential physics of the TAMR effect can be captured in a simple model which was previously introduced in Ref.~\cite{Neel2013} and is summarised briefly in the next section before being used to explain the origin of the TAMR effect for the case of the Ir adatom.

\subsubsection{Model of the TAMR}
The TAMR effect can be modelled using two localised atomic states at a surface interacting via the spin-orbit interaction:
$$
    (E \cdot \dblone - H - \Sigma) G(E) = \dblone
$$
where $H$ is the Hamiltonian matrix:

\noindent\makebox[\textwidth]{
 \( H = \left( \begin{array}{cc}
\epsilon_1 & -t \\
-t & \epsilon_2 \end{array} \right)\).
}
$\epsilon_1$ and $\epsilon_2$, which describe the energies of the two states, and $t$, the hopping between them, all depend on the spin-quantization axis due to the spin-orbit interaction. 
The diagonal elements of the self energy matrix,

\noindent\makebox[\textwidth]{
 \( \Sigma = \left( \begin{array}{cc}
- i \gamma_1 & 0 \\
0 & - i \gamma_2 \end{array} \right)\),
}
describe the broadening of the peaks induced by hybridization of the atomic states with the surface. 
The LDOS, $n_1(E)$ and $n_2(E)$, of the two states is then given by the diagonal elements, $G_{ii}$, of the Green's function matrix given by
\(
n_i(E) = -\frac{1}{\pi} \mathrm{Im}(G_{ii}(E))
\)
\begin{figure}
\begin{centering}
\includegraphics[width=0.45\linewidth]{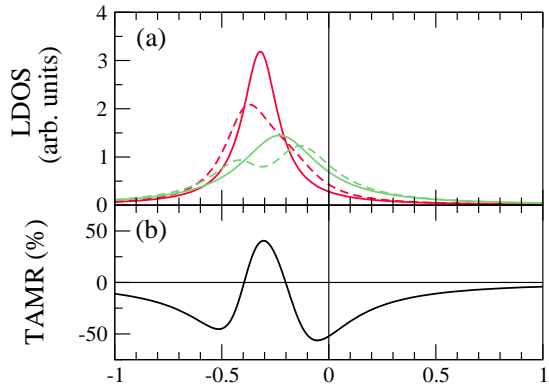}
\caption{\label{fig:model_tamr}(Color online)\,(a) LDOS obtained for the simple model of two atomic surface states that exhibit different orbital symmetry, $d_{z^2}$ (red) and $d_{xz}$ (green), and couple via SOC. The solid (dashed) line refers to the magnetization perpendicular (parallel along [1$\bar{1}$0]) to the film plane. (b) Anisotropy of the LDOS calculation according to Equation.~\ref{eqn:tamr} for the $d_{z^2}$ state in (a).}
\end{centering}
\end{figure}
The broadening of the peaks, $\gamma_1$ and $\gamma_2$, as well as their positions are extracted from the relevant LDOS of the Ir atom in Fig.~\ref{fig:ir_adatom}.
Here we consider the changes induced in the LDOS of the  $d_{z^2}$ and  $d_{xz}$ states at approximately $-0.35$~eV due to changes in the magnetization direction. Their broadening ($\gamma_1 = 0.1$, $\gamma_2 = 0.22$) as well as their energy difference ($\epsilon_1 - \epsilon_2 = -0.09$~eV) are chosen accordingly. 
The mixing of the majority $d_{z^2}$ and $d_{xz}$ states is given by the matrix element \cite{PhysRev.140.A1303}:
$$
|\langle \uparrow, d_{z^2} | H_{SOC} | d_{xz}, \uparrow \rangle | = \nicefrac{1}{2} \sqrt{3} \sin \theta \sin \phi
$$
where $H_{SOC} = \xi \bf{L} \cdot \bf{S}$ is the Hamiltonian of SOC,  $\xi$ is the spin-orbit coupling constant, $\bf{L}$ is the angular momentum and $\bf{S}$ is the spin. This vanishes for an out-of-plane magnetization ($\phi = 90^\circ$, $\theta = 0^\circ$) and is maximal for an in-plane magnetization ($\phi = 90^\circ$, $\theta = 90^\circ$) along the [1$\bar{1}$0] direction. The strength of spin-orbit coupling for 5$d$-transition metals is on the order of 100 -- 500~meV \cite{PhysRevB.13.2433}. The mixing parameter, $t$, is therefore chosen to be 0~meV for the out-of-plane magnetization direction and 150~meV for the in-plane magnetization. Fig.~\ref{fig:model_tamr} shows the result of the model for the case of a Ir adatom adsorbed on the Mn/W(110) surface. 
The LDOS of the  $d_{z^2}$ and  $d_{xz}$ states are enhanced for the out-of-plane magnetization and exhibit a small energy shift in the position of the peak with respect to the LDOS of the in-plane magnetization. This is in agreement with the results of the DFT calculation shown in Fig.~\ref{fig:ir_adatom}(c) and (d). The anisotropy of the LDOS is shown in Fig.~\ref{fig:model_tamr}(b) and is obtained by taking only the  $d_{z^2}$ states into account as these states survive longer in the vacuum.
The resulting shape is similar to the TAMR calculated from the DFT in an energy window between $-0.55$~eV and $-0.32$~eV. Outside this energy region several more states come into play and this model cannot hope to reproduce their interactions. 
Finally, we note that spin orbit coupling will contribute a $\cos^2\theta$ angular dependence to the LDOS, and hence the tunnelling current, due to the shape of the matrix elements determined the mixing between the $d$-states  \cite{PhysRev.140.A1303} (assuming no rotation in-plane, i.e., $\phi = 0$). This will be in addition to the standard $\cos \theta$  angular dependence of the spin-polarised part of the tunnelling current (i.e., without SOC). 

\subsubsection{Spin Polarization}
\begin{figure}[ht]
\begin{centering}
\includegraphics[width=0.45\linewidth]{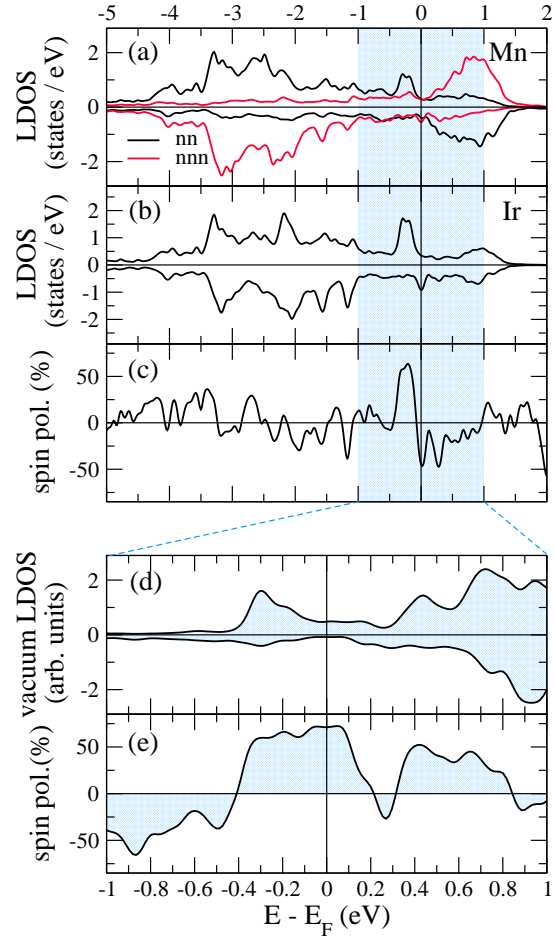}
\caption{\label{fig:ir_spin_pol}(Color online)\,Spin-resolved LDOS of Mn atoms in the Mn ML on W(110) (both nearest neighbour (nn) and next nearest neighbour (nnn) with respect to the Ir adatom). 
(b) Spin-dependent LDOS of the Ir adatom. (c) Spin polarization of the Ir adatom calculated according to Equation.~(\ref{polarization}).  
(d) LDOS in the vacuum 6~\AA\ above the Ir adatom. (e) Spin polarization of the vacuum LDOS presented in (d).}
\end{centering}
\end{figure}

As for the Rh adatom, the hybridization between the Mn atoms of the surface and the Ir adatom leads to a spin polarization of the latter. This can be seen by comparing the spin-resolved LDOS of the Ir atom and its neighbouring Mn atoms as shown in Fig.~\ref{fig:ir_spin_pol}(a) and (b). The strong hybridization is evident in the nearest neighbour Mn majority $3d$-states and Ir $5d$-states at $-0.25$~eV as well as in the minority spin channel at $-3.1$~eV, $-1.2$~eV and at the Fermi level. The reduced exchange splitting of the Mn atoms  compared to Mn on a clean surface results in a smaller magnetic moment in these atoms after the deposition of the Ir adatom compared to the clean surface: $+2.8$~$\mu_B$ compared to $+3.5$~$\mu_B$ (see Table~\ref{tab:magnetic_moments}). As was stated earlier, the small exchange splitting of the Ir states leads to an induced magnetic moment of $+0.2$~$\mu_B$. 
Despite the small magnetic moment, the spin polarization of the adatom can be large. This is shown in Fig.~\ref{fig:ir_spin_pol}(c). For example, the hybridization-induced presence of the majority $d_{z^2}$ state at $-0.25$~eV results in a very large spin polarization of $\approx$~60\%. In contrast, the dominant minority states at the Fermi level have $d_{yz}$ and $d_{xy}$ symmetry resulting in a large negative spin polarization of $-45$\% at the Ir adatom. 
Using a magnetic STM tip, one can then be sensitive to the spin polarization in the vacuum. In Fig.~\ref{fig:ir_spin_pol}(e) we show the spin polarization in the vacuum defined as: 
\begin{equation}\label{polarization}
P(z,\epsilon)=\frac{n^\uparrow(z,\epsilon)-n^\downarrow(z,\epsilon)}{n^\uparrow(z,\epsilon)+n^\downarrow(z,\epsilon)}
\end{equation} where $n^{\uparrow(\downarrow)}(z,\epsilon)$ is the spin-resolved LDOS calculated in the vacuum at a distance, $z$, from the surface.
At the Fermi level it is clear that there is a change in sign in spin polarization as one moves away from the surface, going from $-45$\% at the Ir adatom (Fig.~\ref{fig:ir_spin_pol}(c)) to $+71$\% at a height of 6~\AA\ above the Ir atom (Fig.~\ref{fig:ir_spin_pol}(e)). 
As a comparison, the spin polarization of Rh reaches 62\% in the vacuum at this height while the Co adatom reaches 51\%. This sign change, between the surface and the vacuum, can be explained by considering the orbital decomposition of the Ir atom as shown in Fig.~\ref{fig:ir_adatom}(c) and (d). As for the Co and Rh adatoms, the density of states at the Fermi level is dominated by minority states with $d_{yz}$ and $d_{xy}$ symmetry while the density of majority states is smaller and comprised mainly of $d_{z^2}$ states. Close to the adatom, it is the minority states that generate the negative spin polarization. 
However, these states decay very quickly in the vacuum compared to states with  $d_{z^2}$ character. Far from the surface, states with majority $s$, $p_z$ and $d_{z^2}$ symmetry all contribute to the vacuum LDOS and the spin polarization. In contrast, for the case of Co it is primarily the majority $s$ states which dominate in the vacuum \cite{Serrate2010}. This difference can be explained by the significantly smaller spin-splitting of the Ir atom $d$-states compared to that of the Co $d$-states, leading to a much higher density of Ir $d_{z^2}$ states at the Fermi level.  
This modification of the substrate spin polarization due to a non-magnetic adsorbent was shown previously to occur for the case of small molecules adsorbed on the same surface \cite{PhysRevB.88.155403}.
\begin{figure}[ht]
\begin{centering}
\includegraphics[width=0.45\linewidth]{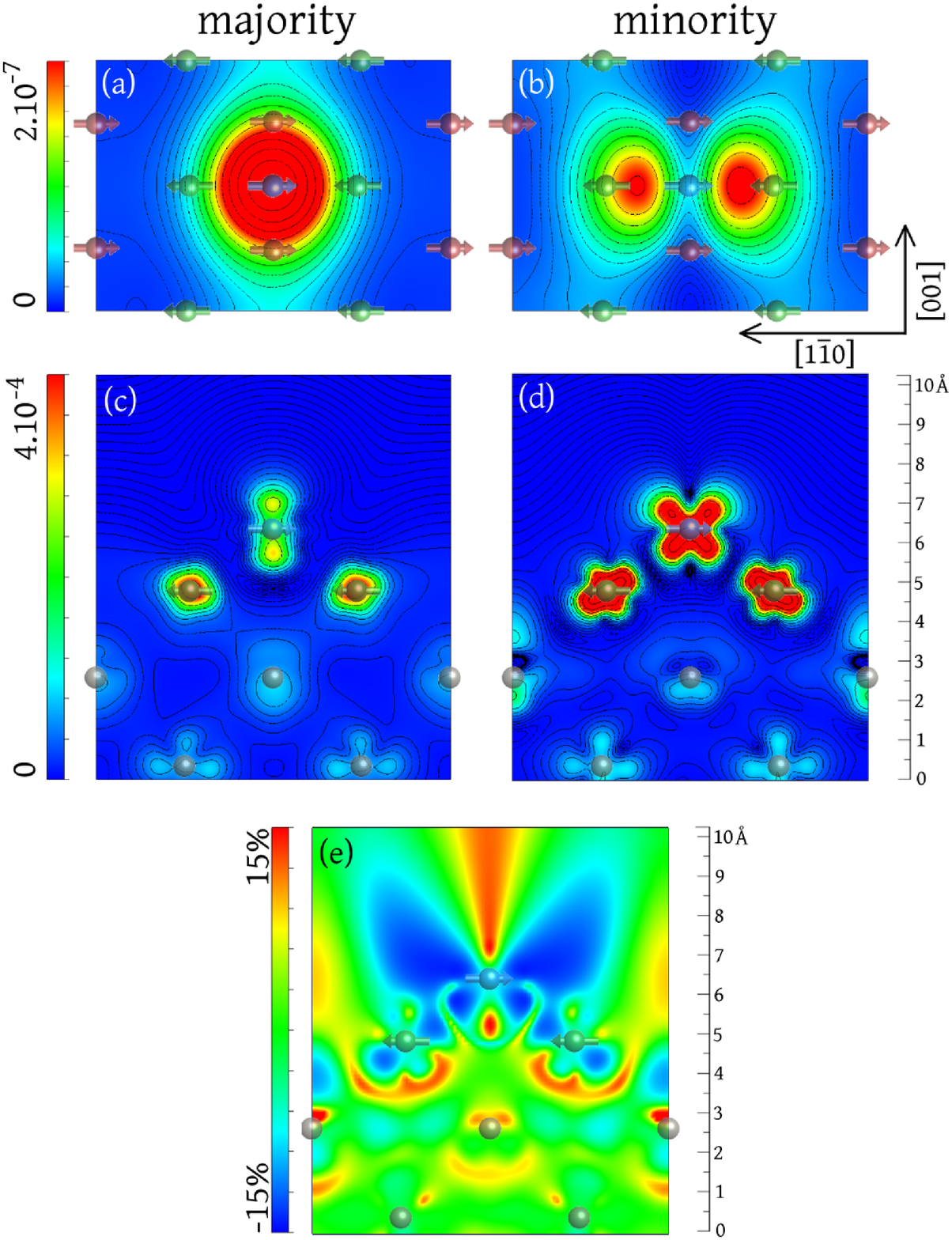}
\caption{\label{fig:stm_images}\,(Color online) 
Spin-resolved partial charge density in the vacuum 3.0~\AA\ above an Ir adatom adsorbed onto a Mn/W(110) substrate, calculated for occupied energy levels in the energy range of [$-10$~meV, $E_F$]. (a) shows the majority contribution to the charge density, (b) the minority contribution, (c) a slice of the majority charge density through the Ir adatom and perpendicular to the surface plane along [1$\bar{1}$0] and (d) a slice of the minority charge density through the Ir adatom and perpendicular to the surface plane along [1$\bar{1}$0]. (e) Slice of the spin polarization calculated according to Eqn.~\ref{polarization} for the data presented in (c) and (d).}
\end{centering}
\end{figure}
In order to visualise these orbital symmetries, Fig.~\ref{fig:stm_images} shows the spin-resolved local density of states integrated in a small energy window below the Fermi energy for the Ir atom adsorbed on the Mn/W(110) surface. 
By using a magnetic STM tip it is possible to image both spin channels separately as demonstrated in Ref.~\cite{Serrate2010} for Co adatoms.
This allows one to view the rotationally symmetric orbitals available in the majority spin channel (Fig.~\ref{fig:stm_images}(a) and (c)), while a double-lobed structure, corresponding to the $d_{yz}$ state, is visible in the minority spin channel (Fig.~\ref{fig:stm_images}(b) and (d)). However, the magnitude of the minority charge density is considerably smaller than that of the majority charge. Therefore, when using a non-magnetic tip, which results in an averaging of both spin channels, only the majority spin channel will be visible. A similar picture emerges for the case of the Rh adatom and was previously shown for the case of the Co adatom \cite{Serrate2010}. Fig.~\ref{fig:stm_images}(c) and (d) show slices of the majority and minority charge density perpendicular to the surface along the [1$\bar{1}$0] direction. Finally, Fig.~\ref{fig:stm_images}(e) shows a slice of the spin polarization determined according to Eqn.~\ref{polarization}, where the spatial distribution can be seen more clearly. Directly above the adatom, the spin polarization changes sign quickly; by approximately 1~\AA\ above the adatom it has reversed sign. Directly above one of the nodes of the minority $d_{yz}$ orbital, however, the spin polarization remains negative for a larger distance above the adatom. This inversion of spin polarisation with distance from the surface was shown previously to occur above Fe adatoms on an Fe(001) surface \cite{ferriani2010origin}.

\section{Summary and Conclusions}
In summary, we have calculated from first-principles the tunnelling anisotropic magnetoresistance effect above single transition-metal atoms adsorbed on a monolayer of Mn on W(110). The spin spiral ground state of this surface provides a magnetic template for adatoms, such that their spins can adopt any of the spin directions offered by the spiral. Here, we have calculated explicitly the two limiting cases - when the magnetization direction points out-of-plane or in-plane along the propagating direction of the spiral. We find a large anisotropic magnetoresistive effect that increases with the nuclear charge of the adatom, rising to a maximum value of approximately 50\% for the case of an adsorbed Ir adatom at $0.35$~eV below the Fermi level. We use a simple model to show the TAMR effect at this atomic limit is driven by magnetization direction dependent mixing of the adatom orbitals.
In addition, we find that the 4$d$ and 5$d$ adatoms experience a spin splitting, induced by their hybridization with the surface, that leads to large spin polarizations in the vacuum. 
Our results demonstrate that the deposition of transition-metal adatoms can be used to determine whether or not noncollinear magnetic ordering is present on magnetic surfaces and that the magnitude of this effect can be increased dramatically using 5$d$ adatoms. To date, studies of the TAMR of single adatoms using an STM tip have been performed by exploiting the domain and domain wall structures of ferromagnetic surfaces. We propose that the use of a spin-spiral surface removes, to a great degree, the constraint imposed by the substrate magnetization direction, and opens up the possibility to study the angular dependence and the magnitude of the TAMR effect on individual adatoms. We emphasise that this effect will be seen even with a non-magnetic STM tip.

\ack
This work was supported by Deutsche Forschungsgemeinschaft via the project B10 of the SFB 677. Computational facilities were provided by the North-German Supercomputing Alliance (HLRN).

\section*{References}



\begin{thebibliography}{10}
\expandafter\ifx\csname url\endcsname\relax
  \def\url#1{{\tt #1}}\fi
\expandafter\ifx\csname urlprefix\endcsname\relax\def\urlprefix{URL }\fi
\providecommand{\eprint}[2][]{\url{#2}}

\bibitem{Julliere1975225}
Julliere M 1975 {\em Physics Letters A\/} {\bf 54} 225

\bibitem{parkin2004giant}
Parkin S~S, Kaiser C, Panchula A, Rice P~M, Hughes B, Samant M and Yang S~H
  2004 {\em Nature Materials\/} {\bf 3} 862

\bibitem{lee2007effect}
Lee Y, Hayakawa J, Ikeda S, Matsukura F and Ohno H 2007 {\em Applied Physics
  Letters\/} {\bf 90} 212507

\bibitem{PhysRevLett.99.067202}
Yayon Y, Brar V~W, Senapati L, Erwin S~C and Crommie M~F 2007 {\em Physical
  Review Letters\/} {\bf 99} 067202

\bibitem{meier2008revealing}
Meier F, Zhou L, Wiebe J and Wiesendanger R 2008 {\em Science\/} {\bf 320} 82

\bibitem{PhysRevLett.103.057202}
Tao K, Stepanyuk V~S, Hergert W, Rungger I, Sanvito S and Bruno P 2009 {\em
  Physical Review Letters\/} {\bf 103} 057202

\bibitem{Serrate2010}
Serrate D, Ferriani P, Yoshida Y, Hla S~W, Menzel M, von Bergmann K, Heinze S,
  Kubetzka A and Wiesendanger R {2010} {\em {Nature Nanotechnology}\/} {\bf
  {5}} {350}

\bibitem{loth2010measurement}
Loth S, Etzkorn M, Lutz C~P, Eigler D and Heinrich A~J 2010 {\em Science\/}
  {\bf 329} 1628

\bibitem{loth2010controlling}
Loth S, von Bergmann K, Ternes M, Otte A~F, Lutz C~P and Heinrich A~J 2010 {\em
  Nature Physics\/} {\bf 6} 340

\bibitem{Ziegler2011}
Ziegler M, N\'{e}el N, Lazo C, Ferriani P, Heinze S, Kr\"{o}ger J and Berndt R
  2011 {\em New Journal of Physics\/} {\bf 13} 085011

\bibitem{PhysRevB.86.180406}
Lazo C, N\'eel N, Kr\"oger J, Berndt R and Heinze S 2012 {\em Physical Review
  B\/} {\bf 86} 180406

\bibitem{khajetoorians2013current}
Khajetoorians A~A, Baxevanis B, H{\"u}bner C, Schlenk T, Krause S, Wehling T~O,
  Lounis S, Lichtenstein A, Pfannkuche D, Wiebe J {\em et~al.\/} 2013 {\em
  Science\/} {\bf 339} 55

\bibitem{PhysRevLett.93.117203}
Gould C, R\"uster C, Jungwirth T, Girgis E, Schott G~M, Giraud R, Brunner K,
  Schmidt G and Molenkamp L~W 2004 {\em Physical Review Letters\/} {\bf 93}
  117203

\bibitem{PhysRevLett.89.237205}
Bode M, Heinze S, Kubetzka A, Pietzsch O, Nie X, Bihlmayer G, Bl\"ugel S and
  Wiesendanger R 2002 {\em Physical Review Letters\/} {\bf 89} 237205

\bibitem{PhysRevB.73.024418}
Shick A~B, M\'aca F, Ma\ifmmode~\check{s}\else \v{s}\fi{}ek J and Jungwirth T
  2006 {\em Physical Review B\/} {\bf 73} 024418

\bibitem{PhysRevLett.98.046601}
Chantis A~N, Belashchenko K~D, Tsymbal E~Y and van Schilfgaarde M 2007 {\em
  Physical Review Letters\/} {\bf 98} 046601

\bibitem{PhysRevB.79.155303}
Matos-Abiague A and Fabian J 2009 {\em Physical Review B\/} {\bf 79} 155303

\bibitem{PhysRevB.80.045312}
Matos-Abiague A, Gmitra M and Fabian J 2009 {\em Physical Review B\/} {\bf 80}
  045312

\bibitem{PhysRevLett.99.226602}
Gao L, Jiang X, Yang S~H, Burton J~D, Tsymbal E~Y and Parkin S~S~P 2007 {\em
  Physical Review Letters\/} {\bf 99} 226602

\bibitem{Viret2006}
Viret M, Gabureac M, Ott F, Fermon C, Barreteau C, Autes G and Guirado-Lopez R
  2006 {\em The European Physical Journal B - Condensed Matter and Complex
  Systems\/} {\bf 51} 1

\bibitem{PhysRevB.86.134422}
von Bergmann K, Menzel M, Serrate D, Yoshida Y, Schr\"oder S, Ferriani P,
  Kubetzka A, Wiesendanger R and Heinze S 2012 {\em Physical Review B\/} {\bf
  86} 134422

\bibitem{PhysRevB.81.212409}
Shick A~B, Khmelevskyi S, Mryasov O~N, Wunderlich J and Jungwirth T 2010 {\em
  Physical Review B\/} {\bf 81} 212409

\bibitem{park2011spin}
Park B, Wunderlich J, Marti X, Hol{\`y} V, Kurosaki Y, Yamada M, Yamamoto H,
  Nishide A, Hayakawa J, Takahashi H {\em et~al.\/} 2011 {\em Nature
  Materials\/} {\bf 10} 347

\bibitem{Bode2007}
Bode M, Heide M, von Bergmann K, Ferriani P, Heinze S, Bihlmayer G, Kubetzka A,
  Pietzsch O, Bluegel S and Wiesendanger R {2007} {\em {Nature}\/} {\bf {447}}
  {190}

\bibitem{Neel2013}
N\'{e}el N, Schr\"{o}der S, Ruppelt N, Ferriani P, Kr\"{o}ger J, Berndt R and
  Heinze S 2013 {\em Physical Review Letters\/} {\bf 110} 037202

\bibitem{Kresse1996}
Kresse G and Furthm\"{u}ller J 1996 {\em Physical Review B\/} {\bf 54} 11169

\bibitem{Kresse1999}
Kresse G and Joubert D 1999 {\em Physical Review B\/} {\bf 59} 1758

\bibitem{Perdew1996}
Perdew J~P, Burke K and Ernzerhof M 1996 {\em Physical Review Letters\/} {\bf
  77} 3865

\bibitem{PhysRevB.50.17953}
Bl\"ochl P~E 1994 {\em Physical Review B\/} {\bf 50} 17953

\bibitem{PhysRevB.13.5188}
Monkhorst H~J and Pack J~D 1976 {\em Physical Review B\/} {\bf 13} 5188

\bibitem{PhysRevB.62.11556}
Hobbs D, Kresse G and Hafner J 2000 {\em Physical Review B\/} {\bf 62} 11556

\bibitem{bode1999growth}
Bode M, Hennefarth M, Haude D, Getzlaff M and Wiesendanger R 1999 {\em Surface
  Science\/} {\bf 432} 8

\bibitem{PhysRevB.31.805}
Tersoff J and Hamann D~R 1985 {\em Physical Review B\/} {\bf 31} 805

\bibitem{PhysRevLett.86.4132}
Wortmann D, Heinze S, Kurz P, Bihlmayer G and Bl\"ugel S 2001 {\em Physical
  Review Letters\/} {\bf 86} 4132

\bibitem{PhysRevB.83.214410}
Palot\'as K, Hofer W~A and Szunyogh L 2011 {\em Phys. Rev. B\/} {\bf 83} 214410

\bibitem{PhysRev.140.A1303}
Abate E and Asdente M 1965 {\em Physical Review\/} {\bf 140} A1303

\bibitem{PhysRevB.13.2433}
Mattheiss L~F 1976 {\em Phys. Rev. B\/} {\bf 13} 2433

\bibitem{PhysRevB.88.155403}
Caffrey N~M, Ferriani P, Marocchi S and Heinze S 2013 {\em Physical Review B\/}
  {\bf 88} 155403

\bibitem{ferriani2010origin}
Ferriani P, Lazo C and Heinze S 2010 {\em Physical Review B\/} {\bf 82} 054411

\end{thebibliography}

\providecommand{\newblock}{}

\end{document}